\documentclass[conference]{IEEEtran}
\IEEEoverridecommandlockouts
\usepackage{cite}
\usepackage{amsmath,amssymb,amsfonts}
\usepackage{algorithmic}
\usepackage{graphicx}
\usepackage{textcomp}
\usepackage{xcolor}
\usepackage[font=small,skip=0pt]{caption}
\usepackage[linesnumbered,ruled,vlined]{algorithm2e}
\SetKwInput{KwInput}{Input}                
\SetKwInput{KwOutput}{Output}              

\SetCommentSty{mycommfont}

\def\BibTeX{{\rm B\kern-.05em{\sc i\kern-.025em b}\kern-.08em
    T\kern-.1667em\lower.7ex\hbox{E}\kern-.125emX}}
\begin{document}

\title{Head Movement Modeling for Immersive Visualization in VR\\
\thanks{This research was supported by the Research Foundation – Flanders (FWO), IDLab (Ghent University – imec), Flanders Innovation and Entrepreneurship (VLAIO), and the European Union. Moreover, the computational resources (STEVIN Supercomputer Infrastructure) and services used were kindly provided by Ghent University, the Flemish Supercomputer Center (VSC), the Hercules Foundation and the Flemish Government department EWI.}
}
\author{\IEEEauthorblockN{Glenn Van Wallendael, Lucas Liegeois, Julie Artois, Peter Lambert}
\IEEEauthorblockA{\textit{IDLab, Ghent University – imec} \\
\textit{Technologiepark-Zwijnaarde 122, 9052 Gent, Belgium}\\
firstname.lastname@ugent.be, https://media.idlab.ugent.be}
}

\maketitle
\setlength{\textfloatsep}{0pt}
\begin{abstract}
Virtual Reality, and Extended Reality in general, connect the physical body with the virtual world. Movement of our body translates to interactions with this virtual world. Only by moving our head will we see a different perspective. By doing so, the physical restrictions of our body's movement restrict our capabilities virtually. By modelling the capabilities of human movement, render engines can get useful information to pre-cache visual texture information or immersive light information. Such pre-caching becomes vital due to ever increasing realism in virtual environments. This work is the first work to predict the volume in which the head will be positioned in the future based on a data-driven binned-ellipsoid technique. The proposed technique can reduce a 1m\textsuperscript{3} volume to a size of 10cm\textsuperscript{3} with negligible accuracy loss. This volume then provides the render engine with the necessary information to pre-cache visual data.  
\end{abstract}


\section{Introduction}
In Extended Reality (XR), the consumer is immersed in a visual environment mainly represented using a combination of meshes and textures~\cite{TextureStreaming} or sometimes a full plenoptic representation~\cite{SMoE}. 
For improved realism, the amount of visual data necessary to generate realism in the render engine has exploded during the last years, making techniques such as pre-caching a necessity. 
For the render engine to pre-cache visual information, it should be able to predict where the participant will be positioned or looking at~\cite{EyeGaze} in the near future. 
Since people have free will, it is impossible and unnecessary for this application to exactly predict their absolute future positions~\cite{LocomotionPrediction}, i.e., gaze tracking.  
So rather than the exact position, this work is the first work to predict the \textbf{volume} of all possible positions. 
This is also different from saliency research~\cite{VRSaliency} because of the need for immediate predictions rather than an analyzable overview result. 
Finally, it is also different from eye-head coordination \cite{VREyeHead} since such techniques derive the eye gaze from the head movement. 

\section{Proposed Method: Binned Ellipsoid}
\begin{figure}[!t]
\includegraphics[width=0.5\textwidth,trim=0 20.5cm 10cm 1.6cm,clip]{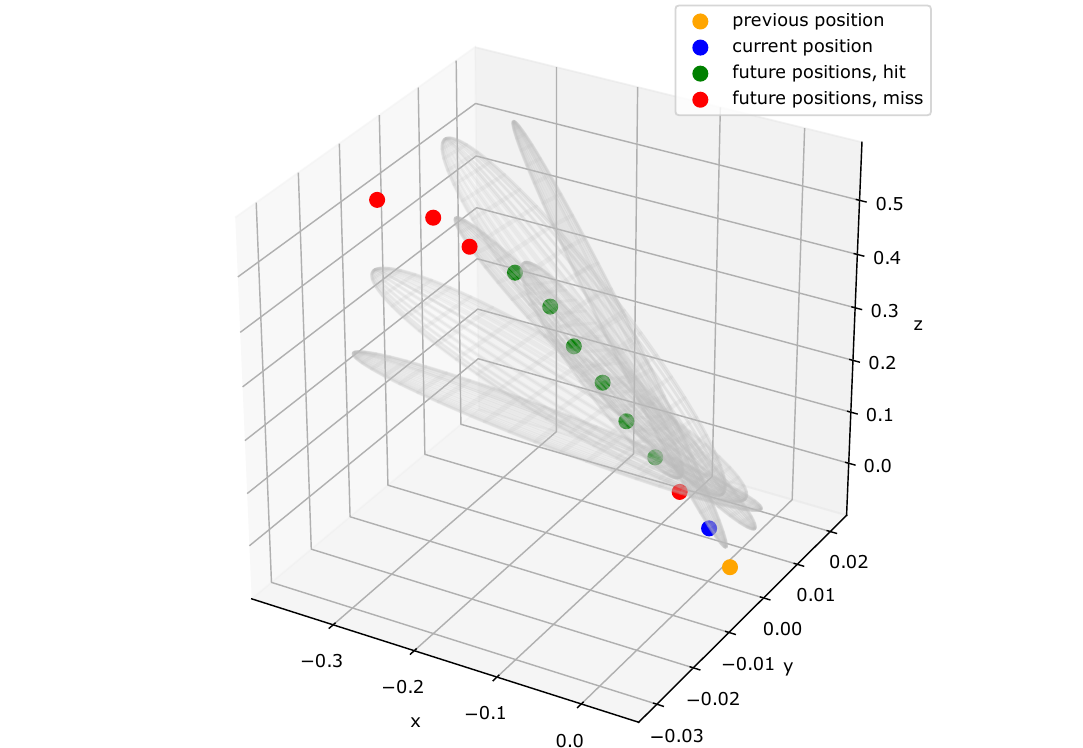}
\caption{The last movement, namely current position minus previous position, corresponds to a bin (Id) which matches to one of these ellipsoids. The volume of the ellipsoid is the prediction resulting in several accurate predictions (green) and several inaccurate predictions (red).}
\label{fig:multipleellipsoid}
\vspace{-0pt}
\end{figure}

This data driven model links every movement made by the headset to an ellipsoid around
its next movements. The model will transform the continuous input space of the known
movements into a discrete space. In practice, this means that every movement will be
mapped to a bin containing similar movements. 
Every bin contains a multitude of possible future positions, all related to near identical preceding movements. 
When queried, the model will calculate
the current movement and look up the corresponding bin. 
As illustrated in Fig.~\ref{fig:multipleellipsoid}, an ellipsoid will be constructed around the union of all future positions, and this will form the prediction. 
Algorithm~\ref{AlgBinEllipsTrain} shows pseudo code for generating predictions with the
proposed model.
\begin{algorithm}[!h]
\DontPrintSemicolon
  \KwInput{$dataset$ (relative positions), $Bsize$ (bin size)}
  \KwOutput{$M$:Bin number (key) - ellipsoids (value) pairs}
  $M \leftarrow empty\ hashmap $ \tcp*{<3D vector, List of ellipsoids>}
  \For{ $Pos\ \textbf{in}\ dataset$}
    {
        $LastMovement \leftarrow Pos_{current} - Pos_{previous}$\\
        $Id \leftarrow LastMovement / Bsize$\\
        \For{ $\ Pos_{next}\ \textbf{in}\ Pos$}
        {
            $NextPosition \leftarrow Pos_{next} - Pos_{current}$\\
            $\textbf{Save}\ Pos,Id,NextPosition$
        }
        
    }
    \For{ $Pos\ \textbf{with}\ \textbf{equal}\ Id$}
    {
        $Ell\ \leftarrow ellipsoid\ around\ NextPositions$\\
        $\textbf{Insert}\ (Id, Ell)\ \textbf{into}\ M$
    }
    $\textbf{return}\ M$
\caption{Training procedure}
\label{AlgBinEllipsTrain}
\end{algorithm}
The bin size decides which movements will be mapped to the same
bin. 
The vector space does not have to be distributed uniformly; the bin sizes can differ per axis.
Head mounted displays often move on a horizontal plane (along the x and z axis) and
move less frequently upwards or downwards (along the y axis). Consequently, the bin size for the y axis can be reduced.

Changing bin size allows for tuning the performance of the model the model. 
On the one hand, bin size has to be small enough,
otherwise all relative movements would be mapped to the same bin. 
The model would not be able to differentiate between the different relative movements, resulting in the same prediction regardless of the input state. 
On the other hand, bin size has to be large enough so that enough relative movements from the training dataset get mapped to the same bin.

Considering computational complexity, the prediction process can be reduced to searching the list of ellipsoids for a given bin number and this requires constant time, O(1), as the data is stored in a hashmap without duplicate keys. 

\section{Results}
Ten different test subjects were tracked, each for a duration of 25 minutes. 
Eight test subjects form the training dataset while the remaining two are used for testing the model.
The test included three movies, namely Special Delivery, Invasion, Rain or Shine, and two games, namely Space Pirate Trainer and Sophie’s Guardian. 
During the movies, the user
typically did not walk much, but they were inclined to look around in the scene because the stories take place in different viewing directions. 
The games, on the other hand, induced rapid and more extreme movements.  

The proposed technique is evaluated on its possibility to predict 10 future samples. For our 90Hz headset, this corresponds to 111ms. 
The possibility to predict such a timeframe in the future is sufficient for local pre-caching operations (from Hard Disk Drive to render engine) and still usable for fast internet-based caching purposes (from cloud to render engine). The accuracy is determined by
counting the percentage of next positions that lie within the
predicted volume.

\begin{figure}[!h]
\includegraphics[width=0.4\textwidth,trim=0.5cm 23cm 11.6cm 1.2cm,clip]{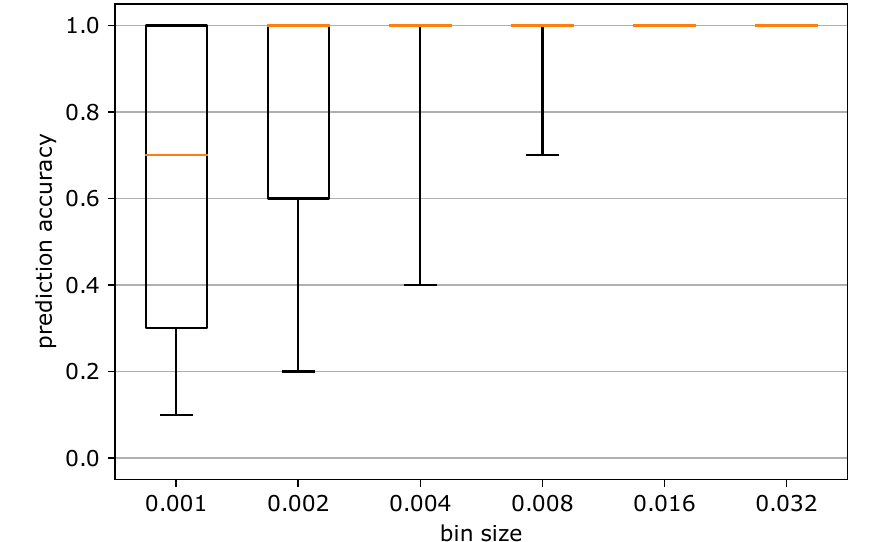}
\caption{Prediction accuracy of the head movement depending on bin size. Closer to 1.0 is better.}
\label{fig:ellipsoidacc}
\end{figure}

\begin{figure}[!h]
\includegraphics[width=0.4\textwidth,trim=0.5cm 22.6cm 11.6cm 1.5cm,clip]{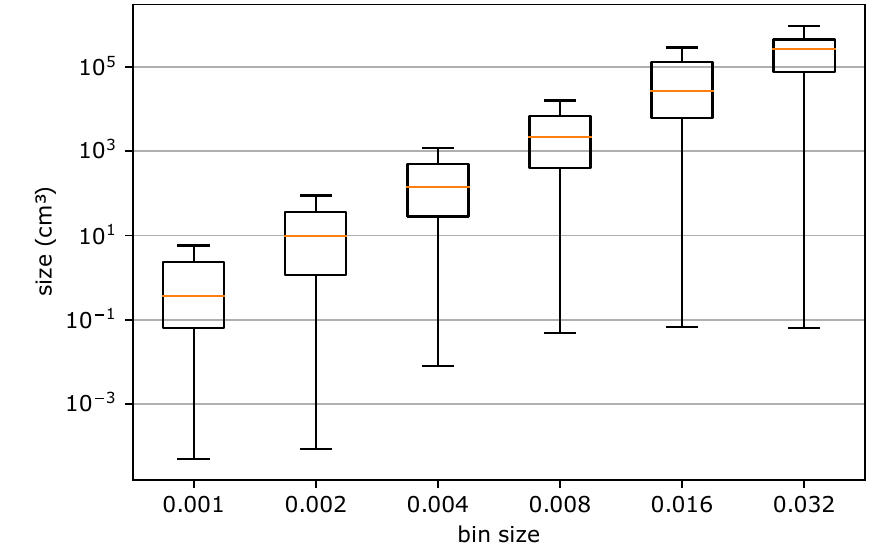}
\caption{Size of the volume in cm\textsuperscript{3} in which the future head position can be expected in function of bin size. Smaller is better.}
\label{fig:ellipsoidvol}
\end{figure}

Fig.~\ref{fig:ellipsoidacc} and~\ref{fig:ellipsoidvol} show the accuracy and the volume of the predictions using various
bin sizes. 
As can be observed in Fig.~\ref{fig:ellipsoidacc}, the bin size can be reduced to 0.002 without significant impact on accuracy. To restrict inaccuracies to outliers only, the bin size should be set to 0.004. 
At these bin sizes, predicted volumes can be reduced from the 1m\textsuperscript{3} total size down to values around 10cm\textsuperscript{3}. 

\section{Conclusion}
The ever-increasing realism in virtual environments requires massive amounts of light information such as textures. 
Render devices are limited in their possibilities and therefore require texture pre-caching techniques. 
Such techniques need an idea about future viewing positions of the player.
In VR environments, such future viewing positions are linked to the physical head movement of the user.
Therefore, predicting head movement is beneficial for the VR render engine. The proposed data-driven technique is based on binned-ellipsoids and is the first technique able to accurately predict the \textbf{volume} in which the head will be positioned in the near future.  
With spatial volumes in the order of magnitude of 10cm\textsuperscript{3}, the position of the head can be predicted 111ms in the future. 
Such timing is enough to pre-cache visual information even from cloud-based services. 
In the future, gaze direction information will help reducing the volume by incorporating viewing-direction specific dimensions.

\bibliography{conference_101719} 
\bibliographystyle{ieeetr}

\vspace{12pt}

\end{document}